\documentclass[9pt]{book}

\usepackage[dvips]{graphicx,color}
\usepackage{makeidx,universe}

\makeauthorindex

\BookTitle{The proceedings of the Physics of Accreting Compact Binaries}

\CopyRight{\copyright 2010 by Universal Academy Press, Inc.}

\begin{document} 

\pagenumbering{arabic}

\chapter{%
Symbiotic Novae}

\author{\raggedright \baselineskip=10pt%
{\bf Joanna Miko{\l}ajewska}\\ 
{\small \it %
N. Copernicus Astronomical Center, Bartycka 18, 00-716 Warsaw, Poland
}
}


\AuthorContents{Joanna Miko{\l}ajewska} 

\AuthorIndex{Miko{\l}ajewska}{j.} 

     \baselineskip=10pt
     \parindent=10pt

\section*{Abstract} 

The symbiotic novae are thermonuclear novae in symbiotic binary systems -- 
interacting binaries with evolved red giant donors, and the longest orbital  periods.  
This paper aims at presenting physical characteristics of these objects and discussing their place among the whole family of symbiotic stars.

\section{Introduction} 

Symbiotic stars are interacting binaries involving an evolved giant (either a normal red giant in S-types or a Mira variable surrounded by an optically thick dust shell in D-types) transferring mass to a hot and compact companion, usually white dwarf. A typical symbiotic binary is embedded in a circumstellar nebula which is mainly formed from material lost in the red giant, while the hot component is responsible for its ionization.
The nature of the giant determines the orbital separation at which the symbiotic interaction occurs: the binary must have enough room for the red giant (and in the case of D-types also for its dust shell) and yet allow it to transfer sufficient mass to its companion. As a result, orbital periods for the S-types are of about 1--15 years, and more than 20 years for the D-types, which are the longest orbital periods among interacting binaries. 
Symbiotic stars are important tracers of late phases of stellar evolution, and a very attractive laboratory to study various aspects of interactions and evolution in binary systems \cite{lapalma02}.
The presence of both the accreting white dwarf and the red giant (with its degenerated core) makes also symbiotic stars a promising ``factory" of Type Ia supernovae (SN Ia), whatever the path a thermonuclear explosion of CO white dwarf upon crossing the Chandrasekhar limit (the single degenerate (SD) scenario) or by merging of a double CO white dwarf system (the double degenerate (DD) scenario) may be.

Symbiotic novae form a small subclass of thermonuclear (TNR) novae which occur in symbiotic binary systems. The aims of this paper are to present the state-of-the-art in understanding of these objects, and to discuss their place among symbiotic stars.

\section{The Nature of Symbiotic Hot Components and their Activity} 

The hot components of symbiotic binaries were discussed by Miko{\l}ajewska \cite{jmik2003,jmik2007,jmik2008}. Based on their activity, all symbiotic stars can be split into two subclasses: ordinary or classical symbiotic stars (majority) and symbiotic novae. 

\begin{figure}[t]
  \begin{center}
  \includegraphics[width=11cm]{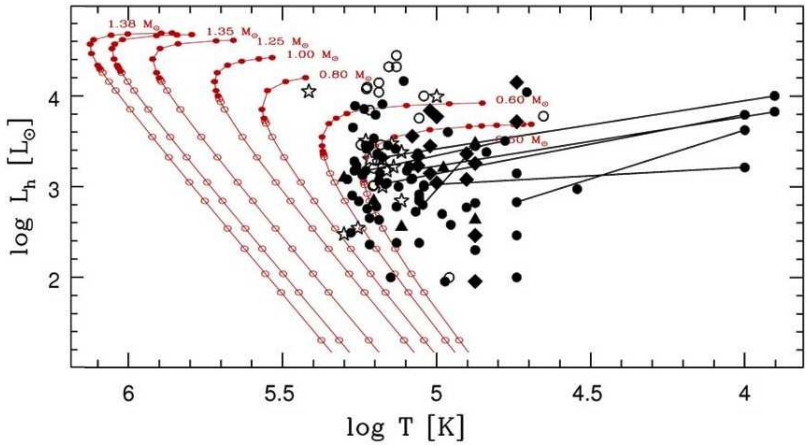}
  \end{center}
  \vspace{-1pc}
  \caption{The hot components in the HR diagram. The luminosities, and effective temperatures are derived from optical and UV emission line fluxes (He{\sc II} and Balmer H{\sc I}) and UV continuum, assuming case B recombination and blackbody energy distribution. Filled and open circles represent the S and D-type systems, respectively (Miko{\l}ajewska, in preparation). Diamonds correspond to measurements by M{\"u}rset et al. \cite{mnsv} whereas stars are data from \cite{jmik2004}, and references therein. Lines connect the points corresponding to Z And-like systems during quiescence and outbursts, respectively.
The solid curves with dots are steady models of Nomoto et al. \cite{nomoto}.}
\label{fig:JMik_HR}
\end{figure}

The typical (quiescent) hot components of classical symbiotics appear to be quite hot ($\sim 10^5$\, K) and luminous ($\sim 100$--$10\,000\,  \rm L_{\odot}$), and they overlap in the same region in the HR diagram as central stars of planetary nebulae (\cite{mnsv, jmik2003}, and Fig.\ref{fig:JMik_HR}). In most cases, the luminosity is too high to be powered solely by accretion. However, if the symbiotic white dwarfs burn hydrogen-rich material as they accrete it, an accretion rate of the order of $10^{-8}\, \rm M_{\odot}/yr$ is then sufficient to power the typical symbiotic component with $M_{\rm WD} \sim 0.5\, \rm M_{\odot}$ \cite{jmik2003}. 

Many of the classical symbiotic systems, including the prototype Z And, CI Cyg and AX Per, show occasional 1--3 mag optical/UV outbursts on timescales from months to years, when the hot component luminosity remains roughly constant whereas its effective temperature varies from $\sim 10^5$ to $\sim 10^4$\, K (Fig.\ref{fig:JMik_HR}). The mechanism of this activity was a mystery for a long time, and only recently it has been qualitatively explained by unstable disc-accretion onto H-shell burning white dwarf \cite{jmik2003,sokol2006a}. Unfortunately, it is still waiting for a good quantitative model. 

The theory predicts that the luminosity of TNR-powered white dwarf should be related to its mass. Unfortunately, there is no unique core mass--luminosity (ML) relation for accreting white dwarfs because it depends significantly on their thermal history: generally, hot white dwarfs should have larger masses than the cooler ones to reach the same luminosity during the H-burning phase (\cite{ib96}). The positions of symbiotic white dwarfs in the luminosity versus mass plane together with different ML relations are shown in Fig.\ref{fig:JMik_hotpar}
The symbiotic white dwarfs cluster around the ML relations for stars leaving the AGB with the CO core and the RGB with the He core for the first time, and this  may indicate that they could be still hot at the onset of the mass transfer from the red giant. The luminosities of the white dwarfs with masses $\geq 0.5\, \rm M_{\odot}$ are generally consistent with those predicted by the steady models of Nomoto et al. \cite{nomoto}, however these models predict systematically higher effective temperatures than those derived from observed UV continuum and UV/optical emission lines (see Fig.\ref{fig:JMik_HR}). 

\begin{figure}[htb]

 \begin{tabular}{cc}
  \begin{minipage}{.45\hsize}
   \begin{center}
     \includegraphics[width=.9\textwidth]{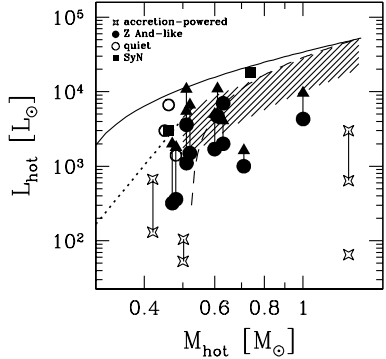}
         \label{fig:JMik_hotpar}
   \end{center}
  \end{minipage}

  \begin{minipage}{.5\hsize}
   \begin{center}
      \caption{Mass--luminosity relation for the symbiotic white dwarfs. Systems with Z And-type activity are represented by filled circles with arrows denoting their luminosity changes. Non-eruptive systems and SyNe during the plateau phase are plotted as open circles, and squares, respectively. Accretion-powered systems (stars) including the SyRNe at quiescence are also shown. 
The ML relation for accreting cold white dwarfs, the Paczy{\'n}ski--Uus relation for ABG stars, and that for He white dwarfs \cite{ib96} are plotted as solid, dashed, and dotted curves, respectively. The shaded region represents 
the steady models from \cite{nomoto}.}
   \end{center}
  \end{minipage}
 \end{tabular}
 \label{fig:JMik_hotpar}
\end{figure}

The symbiotic novae are thermonuclear novae. These are either extremely slow novae, their outbursts can go on for decades, or very fast recurrent novae with very short timescales of their outburst, $\sim$ several days, and recurrence time of order of several years (Fig.\ref{fig:JMik_LC}). The term ``symbiotic nova" (SyNe) is usually reserved for the former whereas the latter are called ``symbiotic recurrent novae" (SyRNe). The differences in outburst behavior seem to reflect different white dwarf masses: very high in the SyRNe, and much lower in the SyNe.

Such novae are very rare among symbiotic stars. Although over 200 symbiotic stars are known \cite{kbel2000},
there are only nine SyNe occurring in both  S-type systems (AG Peg, RT Ser, V1329 Cyg, and PU Vul) and D-type systems (RR Tel, V2110 Oph, V1016 Cyg, HM Sge, and RX Pup), and only four SyRNe (RS Oph, T CrB, V3890 Sgr, and V745 Sco), all of them related to the S-types. 
The most recent symbiotic nova V407 Cyg (2010) shares characteristics of both groups: it has a Mira companion but its very fast outburst development is typical for the SyRNe.

\section{Binary Parameters and Mass Transfer Mode}

\begin{figure}[htb]

 \begin{tabular}{cc}
  \begin{minipage}{.6\hsize}
   \begin{center}
     \includegraphics[width=.9\textwidth]{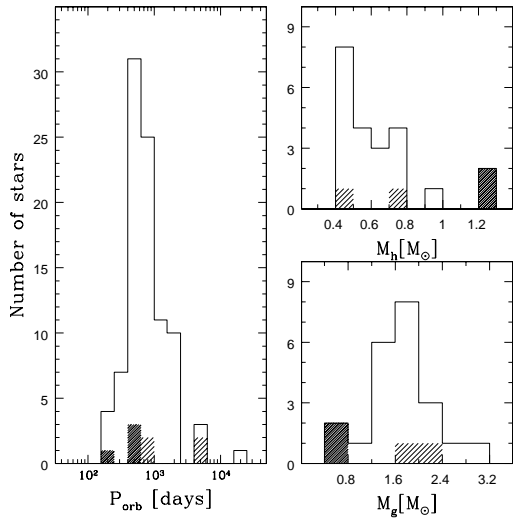}
         \label{fig:JMik_par}
   \end{center}
  \end{minipage}

  \begin{minipage}{.3\hsize}
   \begin{center}
     \vspace{4cm}
      \caption{Orbital parameters of symbiotic stars. Shaded regions denote locations of the SyNe (lighter) and the SyRNe (darker), respectively.}
   \end{center}
  \end{minipage}
 \end{tabular}
 \label{fig:JMik_orbpar}
\end{figure}

The distributions of the orbital parameters and the component masses have been recently discussed in \cite{jmik2007,jmik2008}. Since then, new orbital periods \cite{schaf2009,marg2010} and component masses \cite{hinkle2009,gm2009} have been derived.
The updated distributions of the orbital parameters are shown in Fig.\ref{fig:JMik_orbpar}, and they do not affect the previous conclusions.
In particular, all but one (R Aqr) systems with measured orbital periods belong to the S-type, and most of them have $P_{\rm orb} \sim 300$--1000 days. 

At present, the orbital periods are known for all SyRNe, and the four SyNe in S-type systems. The periods of all SyRNe are below $\sim$ 600 days whereas the SyNe have $P_{\rm orb} \geq 800$ days, consistent with the shorter and longer period tail of S-type systems, respectively. In particular, RT Ser and PU Vul with $P_{\rm orb}$ of 4500 and 4900 days, respectively, are among the systems with the longest known periods (the third one is an accretion-powered CH Cyg). 
It is also remarkable that whereas only $\sim 20\%$ of known symbiotic binaries belong to the D-type, six of ten known SyNe have a Mira companion.
This may be related to lower mass transfer and accretion rate in the SyNe than in most of classical S-type symbiotic stars, and the SyRNe.

The masses of the symbiotic red giants peak around $1.6\, \rm M_{\odot}$, and most white dwarfs have masses between 0.4--0.6\, M$_{\odot}$. There are, however, significant differences between the SyRNe and the other symbiotic binaries. In both, RS Oph and T CrB, the giant is the less massive component, with mass of only $0.4-0.8\, \rm M_{\odot}$, whereas their white dwarfs, $\sim 1.1$--$1.4\, \rm M_{\odot}$,  are the most massive, sufficient to become SN Ia. The red giant envelope mass in both SyRNe is very low, less than $\sim 0.35\, \rm M_{\odot}$  in RS Oph, and below $\sim 0.2$ (and maybe even 0.1) $\rm M_{\odot}$ in T CrB. Stripping this envelope by the supernova ejecta would leave a low mass, $< 0.5\, \rm M_{\odot}$, helium white dwarf only. So, the SyRNe with their giant donors can also account for the single low mass white dwarfs.  

It is usually assumed that the symbiotic binary components do not fill their Roche lobes, and interact via stellar wind. However, during the last decade, ellipsoidal variability has been detected in near infrared and red light curves of over a dozen classical symbiotics suggesting that the giants are at least very close to filling their Roche lobes \cite{jmik2007,jmik2008}. Such an ellipsoidal modulation is also present in all SyRNe but RS Oph \cite{jmik2007,schaf2009}. Unfortunately, to see this modulation in classical symbiotic systems we need red/near-IR  photometry which is hardly available. Thus it is possible that tidally distorted donors and Roche-lobe overflow are quite common is symbiotic binaries with $P_{\rm orb}$ below 1000 days, and the mass transfer and accretion is efficient enough to power the hot components of the classical symbiotic systems via stable/quasi-stable H-shell burning.

In the case of the SyRNe, relatively high, of order of $10^{-7}\, \rm M_{\odot}/yr$,  mass accretion rates are required to account for their short recurrence time, and the hot component activity between the TNR nova outbursts \cite{am1999,marg2008}, which is easier to achieve by Roche-lobe overflow than by wind interaction.
On the other hand, the slow SyNe should accrete at lower rates, below $\sim 10^{-8}\, \rm M_{\odot}/yr$, as in the case of wind accretion. This explains the higher occurrences of the SyNe among D-type systems, and S-type systems with the longest orbital periods. 
Moreover, the SyNe AG Peg is the only symbiotic binary with very good near-IR light curve and $P_{\rm orb} < 1000$ days which does not show any evidence for ellipsoidal variability \cite{jmik2008}.

\section{Outburst Evolution of Symbiotic Novae}

\begin{figure}[htb]
  \begin{center}
    \includegraphics[height=4 cm]{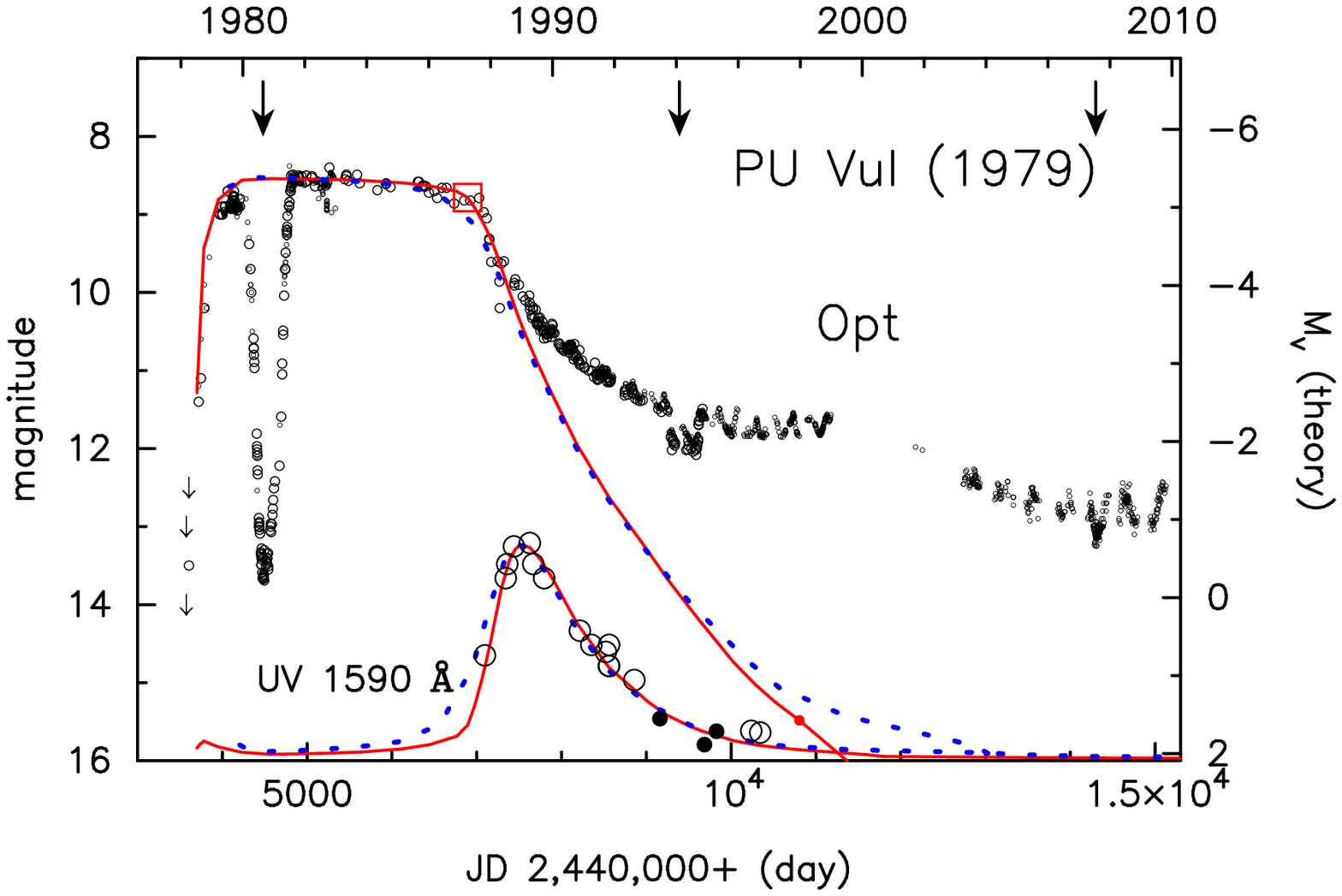}
 \hspace*{0.3cm}%
 \includegraphics[height=4 cm]{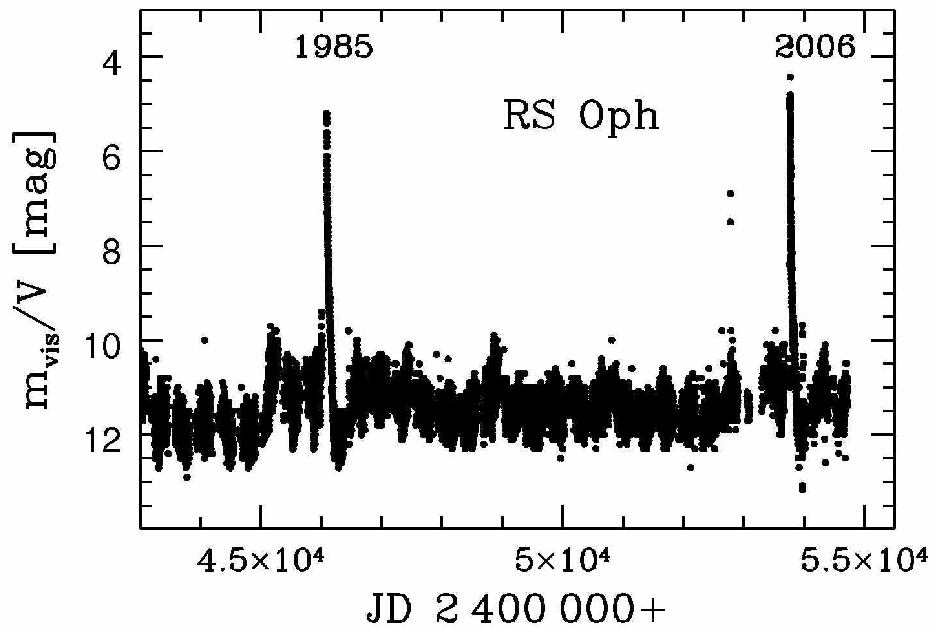}
  \end{center}
  \vspace{-1pc}
  \caption{Optical and UV light curves of the symbiotic nova PU Vul (left) and the AAVSO optical light curve of the symbiotic recurrent nova RS Oph (right) covering identical time intervals.
For PU Vul, the best fit model with $0.6\, \rm M_{\odot}$ white dwarf \cite{kato2010b} is also shown.}
\label{fig:JMik_LC}
\end{figure}

Fig.\ref{fig:JMik_LC} presents light curves from the ongoing outburst of the symbiotic nova PU Vul (1979), and of the two recent outbursts of the symbiotic recurrent nova RS Oph, respectively. Both light curves cover the same interval of time to show dramatic differences in the outburst development between the two kinds of symbiotic novae. 

The very fast outburst evolution of RS Oph, and other SyRNe is due to the presence of the very massive white dwarf (e.g. $\sim 1.3\, \rm M_{\odot} in RS Oph $\cite{brandi2009}) accreting at a rate of order of $10^{-7}\, \rm M_{\odot}/yr$.
The luminosity of the nova at maximum is super-Eddington, and the constant-luminosity phase (plateau) lasts $\sim 50$ days only. The outburst material is ejected with a high initial velocity, $v_{\rm exp} \geq 4000\, \rm km/s$. Relatively low mass of the nova ejecta, $\sim 10^{-7} M_{\odot}$ in RS Oph \cite{sokol2006b}, comparable to the average accretion rate between the TNR nova outbursts, suggests that the white dwarf mass can grow.
The outburst behavior of RS Oph, and the recurrent nova phenomenon in general, has been recently discussed in \cite{keele2008}. 

The outbursts of PU Vul and other SyNe develop very slowly: the rise to maximum takes months to years, and their decline to the pre-outburst state can take decades. The outburst of PU Vul began in late 1977, and while the optical maximum was reached in 1979, the system has not yet recovered to its pre-outburst stage. The record-holder among the SyNe is however AG Peg: the constant-luminosity phase lasted for about 120 years, and the plateau luminosity, $\sim  3500 \pm 1000\, \rm L_{\odot}$, was the lowest among the whole sample (see Fig.6 in \cite{jmik2003}). Most of the SyNe develop an A/F-type spectrum at optical maximum, similar to that of classical novae; however without large violet displacements of the absorption lines which may indicate lack of massive optically thick wind \cite{kato2010a}. In V1016 Cyg, V1329 Cyg, and HM Sge such a spectrum was not detected, possibly because of a lack of early enough spectroscopic observations. 

\begin{figure}[htp]
  \begin{center}
    \includegraphics[width=8 cm]{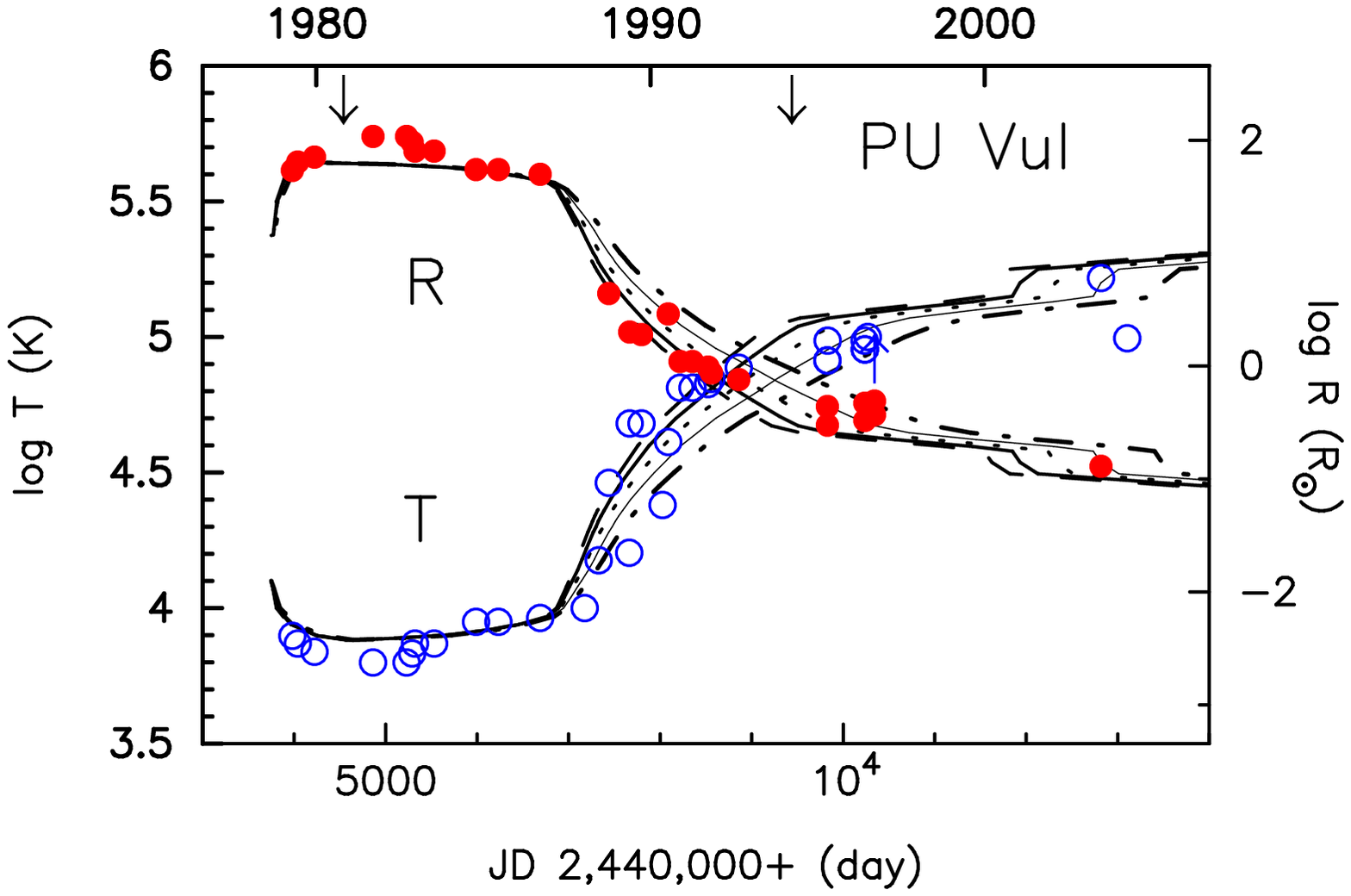}\\
   \vspace*{0.5cm}%
 \includegraphics[width=8 cm]{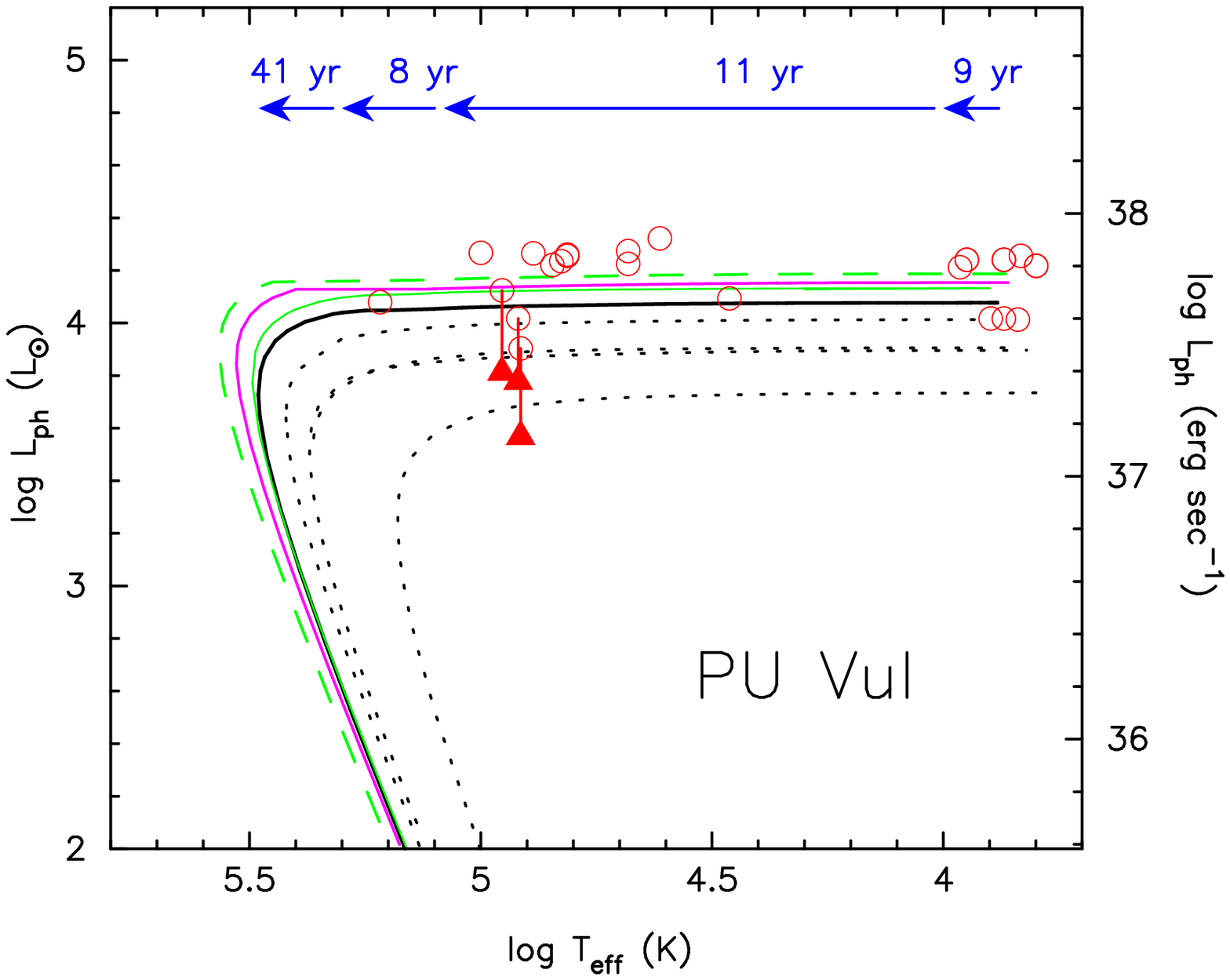}
  \end{center}
  \vspace{-1pc}
  \caption{Top: The evolution of the effective temperature and radius of PU Vul. Lines denote the blackbody photospheres with different mass-loss rates: $4 \times 10^{-7}$ (dash-dotted),   $ 5 \times 10^{-7}$ (thin solid),  $6 \times 10^{-7}$ (dotted),   $7 \times 10^{-7}$ (thick solid), and  $8 \times 10^{-7}$ (dashed), respectively. Bottom: The HR diagram. Lines denote HR tracks for different models. Solid lines:  $0.6\, \rm M_{\odot}$, X=0.7, Z=0.01 (thick black -- the best fit model), $0.6\, \rm M_{\odot}$, X=0.5, Z=0.006 (magenta), $0.7\, \rm M_{\odot}$, X=0.5, Z=0.006 (green), respectively. Dashed green line denotes a $0.7\, \rm M_{\odot}$ with X=0.7, and Z=0.006. Dotted lines represent cold $0.5\, \rm M_{\odot}$ white dwarfs with different X and Z. The lowest dotted line represents a hot $0.5\, \rm M_{\odot}$ white dwarf with X=0.7, and  Z=0.02. Arrows indicate evolution timescales for the best fit model. }
\label{fig:JMik_PUVul}
\end{figure}

The outburst evolution of PU Vul has been recently studied by Kato et al. \cite{kato2010a,kato2010b}. 
Their analysis of the observed optical and UV light curves with quasi-evolution model consisting of a series of static solution resulted in the best fit mass of the white dwarf of $\sim 0.6\, \rm M_{\odot}$ (Fig.\ref{fig:JMik_LC}). Because of the low white dwarf mass, the TNR event is very slow and quiet without massive optically thick wind, the white dwarf can retain most of the accreted mass \cite{kato2010a}.

Fig.\ref{fig:JMik_PUVul} shows development of the effective temperature and radius of the hot component of PU Vul, and its evolution in the HR diagram together with different models from \cite{kato2010b}. The hot component evolves towards higher effective temperatures, and its radius shrinks as the nova declines from optical maximum. These changes are substantial; the effective temperature was $\sim 160\,000$\,K in 2003, while $T_{\rm eff} \sim  6000$--7000\,K at visual maximum. Similarly, the radius decreased from its maximum value of $\sim 100\, \rm R_{\odot}$ in 1982-83 to $\sim  0.13\, \rm R_{\odot}$ in 2003. Despite these temperature and radius changes, the luminosity remained constant at least till mid 2000's, in agreement with the timescales predicted by the best fit model with $0.6\, \rm M_{\odot}$ white dwarf \cite{kato2010b}. 

\begin{figure}[htb]

 \begin{tabular}{cc}
  \begin{minipage}{.55\hsize}
   \begin{center}
     \includegraphics[width=.9\textwidth]{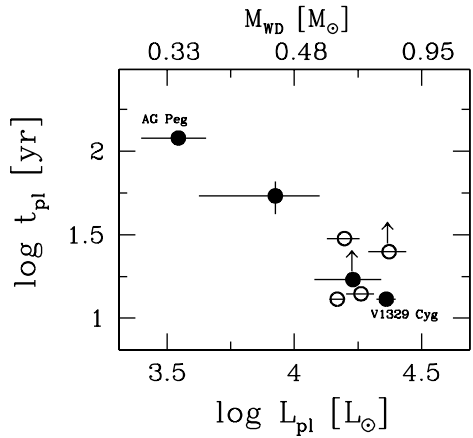}
   \end{center}
  \end{minipage}

  \begin{minipage}{.35\hsize}
   \begin{center}
     \vspace{3cm}
      \caption{The plateau lifetime-luminosity relationship for symbiotic novae. To associate mass and luminosity on the top axis of the figure, the mass-luminosity relationship for the plateau phase from \cite{ib96} was adopted.}
\label{fig:JMik_LTplateau}
   \end{center}
  \end{minipage}
 \end{tabular}
\end{figure}

The outburst behavior of PU Vul is similar to that of other SyNe \cite{mk1992,ken1993,mn1994,jmik1999}, 
although their plateau luminosities and timescales of the outburst significantly differ.  
Fig.\ref{fig:JMik_LTplateau} presents correlation between the lifetime ($t_{\rm pl}$) and luminosity ($L_{\rm pl}$) of SyNe during their plateau phase. Apparently, the lower is the luminosity, the longer time the nova spends on the plateau. Such a behavior is expected from the theoretical studies, and reflects dependence of both parameters on the white dwarf mass. For two of these novae, AG Peg and V1329 Cyg, the white dwarf masses have been derived from radial velocity curves \cite{jmik2003}, and they are in good agreement with the theoretical prediction. AG Peg with its very low,  $0.46+/-0.10\, \rm M_{\odot}$,  mass of the white dwarf is the slowest nova even recorded whereas one of the fastest of the SyNe, V1329 Cyg, contains much more massive, $0.74+/-0.08\, \rm M_{\odot}$, white dwarf.

Interestingly, the presence or absence of the A/F-type spectrum at maximum is not obviously related to the plateau characteristics which may indicate that this is an observational effect rather than a real physical difference. In particular, V1329 Cyg, V1016 Cyg and HM Sge are the most luminous SyNe, and if their A-type phase was shorter than in the other SyNe we could have lost it.
Another possibility is that they underwent non-degenerate shell flashes on hot white dwarfs \cite{kt1983}. 
Unfortunately, there are only few pre-outburst observations for the SyNe \cite{mk1992}. On old objective prism plates, three of them, V1016 Cyg, V1329 Cyg, and PU Vul, look as M giants, and only V1016 Cyg shows a strong H$\alpha$ emission indicating the presence of a hot white dwarf prior to its eruption.
The situation is much better in the case of RX Pup \cite{jmik1999} which have exhibited irregular photometric variability accompanied by significant spectroscopic changes. In particular, in 1941 is showed optical continuum with very strong high-excitation emission lines, whereas its first spectrum taken in 1894, resembled that of $\eta$\,Car. The hot component clearly showed more activity before its 1970 nova outburst than any other symbiotic nova, and its white dwarf should have been the hottest among them. However, its evolution in the HR diagram during its nova outburst is similar to that of PU Vul, with the A/F-type spectrum maintained during 1972-79 \cite{jmik1999}.

\section{The Case of V407 Cyg} 

\begin{figure}[htb]
  \begin{center}
\includegraphics[width=10cm]{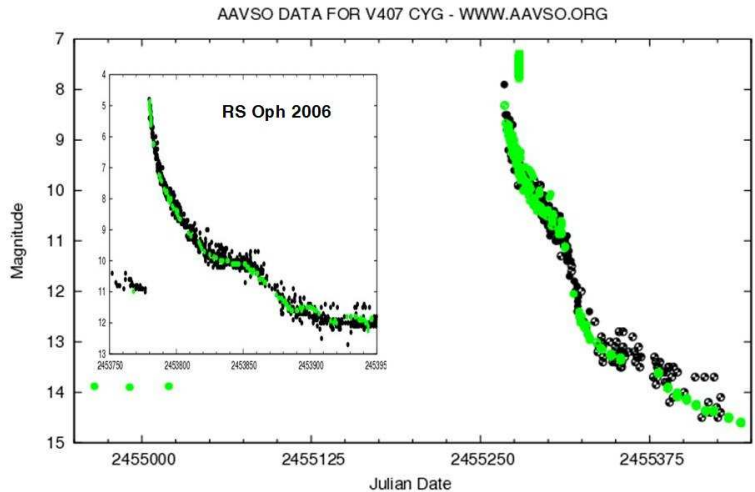}
  \end{center}
  \vspace{-1pc}
  \caption{The AAVSO optical light curve of V407 Cyg compared to that of the 2006 outburst of RS Oph (insert). The horizontal scale (JD) is the same for both light curves.}
\label{fig:JMik_v407cyg}
\end{figure}

The most recent nova eruption of V407 Cyg is both unique and challenging.
The outburst started in the first week of March 2010, and its optical spectrophotometric development is very similar to that of RS Oph. In particular, the very fast rise to maximum and decline within a few months to its pre-outburst visual brightness indicates that the white dwarf is massive as in the SyRNe (Fig.\ref{fig:JMik_v407cyg}). 
It is very surprising because V407 Cyg was hitherto known as a classical D-type symbiotic system which means that it must be wind-accreting system.

The Mira component of V407 Cyg is extreme, in the sense that it has the longest pulsation period among symbiotic Miras ($P_{pul}=763$ days \cite{shugar2007}), and the only one which is Li-rich \cite{tatar2003}.
Such a Li enhancement is very rare among evolved giants -- it has been found in the LMC Miras with very long, $\geq 400$ days, periods, and only in one galactic Mira with a period of $\sim 700$ days -- and it is usually explained as a consequence of hot bottom burning (HBB) \cite{tatar2003}. Since the HBB occurs only in stars with initial masses in the range of 4--8 $\rm 
 M_{\odot}$, the first formed white dwarf component of V407 Cyg should have had a massive progenitor, $ >  4$--$8 M_{\odot}$, so it can be indeed very massive.

There are also other remarkable similarities between V407 Cyg and the SyRNe.
Before the recent TNR outburst, V407 Cyg showed activity (high and low stages) similar to that observed in CH Cyg and other accretion-powered symbiotic stars \cite{shugar2007}, as well as the intrinsic variability of the hot components of the SyRNe between their TNR nova outbursts.  In particular, in all these systems the activity is characterized by very similar timescales, and a variable A/F-type shell source with $L_{\rm UV/opt} \sim 10$--$1000\, \rm L_{\odot}$  accompanied by flickering, similar to that observed in cataclysmic variables,  appears at the bright (high) stage \cite{am1999,marg2008,jmik2003,jmik2008,shugar2007}.
Moreover, the same timescales characterize multiple outburst activity of Z And-type symbiotic stars, and the same correlation between flickering and activity was found in Z And itself during its last series of eruptions \cite{sokol1999}.
The activity of Z And-type symbiotic stars and the high and low states of the SyRNe, CH Cyg, and other similar systems have been associated with unstable disc-accretion onto the white dwarf (e.g. \cite{jmik2008}). 
In the case of V407 Cyg, the hot component luminosity during the active state in 1998, $\sim500 L_{\odot}$ \cite{shugar2007}, requires $\dot{M}_{\rm acc} \sim 10^{-7}\, \rm M_{\odot}/yr$ (the distance of $\sim 2.7$ kpc with the period-luminosity relation was adopted), similar to that expected in the SyRNe. Although such an accretion rate is somewhat high for a wind-accreting system, it is not impossible. Symbiotic Miras are known as sources of strong stellar winds with rates of about of $10^{-6}$--$10^{-5}\, \rm M_{\odot}/yr$, and those with the longest pulsation periods have the highest mass-loss rates \cite{marg2009}. 
Moreover, recent theoretical simulations of interactions and wind-accretion in symbiotic binaries with dusty Mira donors show that once the dust-formation radius compares within a factor of a few to the Roche-lobe radius, one can expect wind Roche-lobe overflow, where most of the wind is trapped within the binary and can efficiently accreted by the companion \cite{mohamed}. It is also very likely that the binary orbit is eccentric, and enhanced mass transfer rate near periastron passage causes a brightening of the accreting component.

Finally, is V407 Cyg the first recurrent nova in D-type symbiotic system?
The answer is NO because by definition at least two TNR explosions must be recorded whereas the 2010 outburst of V407 Cyg is the first one of this type. There have been some suggestions, also during this meeting, that the 1936 brightening was due to another nova outburst, however both amplitude and timescale point to a high state similar to that in 1998, although somewhat fainter and shorter \cite{shugar2007}.

\section{Summary}

Most of the symbiotic hot components are low mass white dwarfs powered by more or less stable thermonuclear burning of the H-rich material accreted from their giant companion. However, they seem to be systematically cooler than the steady models predict.

A thermonuclear nova explosion is a very rare phenomenon in symbiotic stars. There are two kinds of symbiotic novae.
The ordinary SyNe (9 hitherto recorded) are very slow and quiet (with no massive optically thick wind) TNR explosions on wind-accreting low to medium-mass white dwarfs which can retain most of the accreted mass (e.g. \cite{kato2010a,kato2010b}).
The SyRNe are TNR explosions on very massive, $\sim >1.2 M_{\odot}$, white dwarfs accreting at very high rates, $\sim 10^{-7}M_{\odot}/yr$ via Roche-Lobe overflow. Low mass of the nova ejecta, $\sim 10^{-7} M_{\odot}$, in RS Oph indicates that the white dwarf mass can grow, and thus they can be progenitors of the SN Ia.
However, the most recent symbiotic nova, V407 Cyg, does fit well this picture because it seems to harbor a massive, $> \sim 1M_{\odot}$, white dwarf as do the SyRNe. The white dwarf is accreting at relatively high rate, at least occasionally reaching the highest rates in the SyRNe.          

Finally, prior to its nova explosion, V407 Cyg has been recognized as accretion-powered symbiotic system sharing pronounced similarity with CH Cyg and R Aqr \cite{shugar2007}, and even Mira itself \cite{sokol2010}. They are several other symbiotic systems with less luminous hot components which can be accretion-powered. It is very likely, that they all have experienced, and will continue to experience symbiotic nova explosions.

\section{Acknowledgements} 
I am very grateful to Mariko Kato for providing figures presenting the outburst behavior of PU Vul. 
I acknowledge with thanks the variable star observations from the AAVSO International Database contributed by observers worldwide and used in this research.
This study was partly supported by Polish Research Grant No. N203 395534.



\end{document}